\magnification=\magstep1
\hsize=125mm
\font\title=cmbx10 scaled \magstep2
\overfullrule=0pt

                                                     DAMTP/R-96/41
\bigskip

\bigskip

\bigskip

\bigskip

\bigskip

\bigskip

\bigskip

\bigskip

\bigskip

\bigskip

\centerline{\title Black Hole Spectrum: Continuous or Discrete?}

\bigskip

\bigskip

\centerline{Jarmo M\"{a}kel\"{a}\footnote*{e-mail address: 
j.m.makela@damtp.cam.ac.uk}}

\bigskip

\centerline{\it Department of Applied Mathematics and Theoretical
Physics, University of Cambridge,}

\centerline{\it Silver Street, Cambridge CB3 9EW, United Kingdom}

\bigskip

\bigskip

\centerline{\bf Abstract}

\bigskip

     We formulate a qualitative argument, based on Heisenberg's
uncertainty principle, to support the claim that when the effects of matter
fields are assumed to overshadow the effects of quantum mechanics of
spacetime, the discrete spectrum of black hole radiation, as such as 
predicted by Bekenstein's proposal for a discrete black hole area spectrum, 
reduces to Hawking's black-body spectrum.

\vfill\eject

          Among the few well-established predictions
obtained from the study of quantum mechanics linked to
Einstein's general relativity is Hawking's celebrated result that black
holes are not completely black but they emit radiation.[1] This
radiation is essentially thermal:  the black
hole emits field quanta of all frequencies according to the usual
black-body  spectrum, corrected with a factor that accounts for the
scattering from the spacetime curvature. In
particular, for a Schwarzschild black hole with mass $M$ the characteristic
temperature is
$$
T = {{\hbar c^3}\over{8\pi Gk}}\,{1\over M}.\eqno(1)
$$

     However, several authors have raised the possibility that Hawking
radiation might in fact have a {\it discrete\/} spectrum. In its earliest
form, this argument traces back to Bekenstein's 1974 proposal [2] that the
eigenvalues of the black hole event horizon area are of the
form
$$
A_n = \alpha n l_{Pl}^2,\eqno(2)
$$
where $\alpha$ is a constant of order one, $n$ ranges over positive
integers, and
$$
l_{Pl} := \left({{\hbar G}\over{c^3}}\right)^{1\slash 2}
\eqno(3)
$$
is the Planck length. This proposal has since been revived on various
grounds,[3-20] one of the latest being its derivation
for a Schwarzschild black hole from
a Hamiltonian quantum theory of
spherical symmetric, asymptotically flat vacuum spacetimes.[21]
In the particular case of a Schwarzschild hole, the fact that Eq.(2)
implies a discrete spectrum for Hawking
radiation can be seen by
recalling that the area of such a black hole with mass $M$ is
$$
A = {{16\pi G^2}\over{c^4}}\,M^2,\eqno(4)
$$
and so it follows from Eqs.\ (2) and (4) that
the angular frequencies of the quanta of the Hawking
radiation are integer multiples of the fundamental
angular frequency
$$
\omega_0 := \alpha{{c^3}\over{32\pi G}}\,{1\over M}.\eqno(5)
$$
For example, if $M$ is ten solar masses, or
$2\times 10^{31}$~kg,
then
$\omega_0$ is of order $0.1$~kHz,
which is roughly the resolving power
of an ordinary portable radio receiver.

         So, we have these two theories with obviously contradicting
experimental predictions.
Hawking's theory implies a continuous black-body spectrum for black hole
radiation, whereas the theory based on Bekenstein's proposal states
that the spectrum is discrete.
The qualitative difference is clear: one easily sees
from Wien's displacement law that the fundamental angular frequency
$\omega_0$ of Eq.(5) is near the angular frequency
corresponding to the maximum of the black-body spectrum with the
temperature of Eq.(1). The question is therefore: Is there any
relationship between these two theories? In what follows, we shall, for the
sake of convenience, refer to the theories based on Hawking's result
and Bekenstein's proposal, respectively, by
{\it continuous\/} and {\it discrete\/}
theories of black hole radiation.

            To begin with, we note that the starting points of the two
theories are  completely different. The continuous theory is a
semi-classical theory where spacetime is treated classically and matter fields
are treated quantum mechanically,
whereas the theories producing Bekenstein's proposal (2), and
hence the discrete theory, are (in most cases) supposed to be quantum
theories of vacuum black hole spacetimes. Given the different starting
points, one should perhaps not be surprised at the difference in the
predictions.

    Now, if one adopts the viewpoint that Bekenstein proposal (2) arises from a
quantum theory of vacuum spacetimes, one may feel justified to regard the
discrete theory as more fundamental than the continuous one. More
precisely, one may expect the continuous theory to emerge as the
semi-classical limit of the discrete theory when the effects of matter
fields are taken into account. This means that if the effects of quantized
matter fields are assumed to overshadow the effects of quantum
mechanics of the spacetime, the discrete theory should reduce to
the continuous one. The object of this paper is to study whether this is
the case. A rigorous treatment of this problem would involve a
simultaneous quantization of spacetime and matter
fields---a problem that we shall not attempt to discuss here at a
quantitative level. However, we shall formulate a
qualitative argument in favor of the claim that the continuous theory is
the semi-classical limit of the discrete one in the sense described above.

            To proceed, we recall that the continuous theory implies
that the luminosity of a Schwarzshild black hole is proportional to
$M^{-2}$. For example, it has been estimated by Page [22-25] that
for a Schwarzschild black hole of mass
$M\gg 10^{14}$~kg,
the luminosity is
$$
L=3.4\times 10^{33}{{{\rm W}{\rm kg}^2}\over{M^2}},\eqno(6)
$$
when the radiation consists
of 81 per cent neutrinos (four types), 17 per cent photons, and 2 per cent
gravitons.
Naively, the $M^{-2}$-dependence of the
luminosity can be obtained from the luminosity given by the
Stefan-Boltzmann law,
$$
L_{SB} = \sigma A_{eff}T^4,\eqno(7)
$$
where $\sigma$ is the Stefan-Boltzmann constant, by substituting for $T$ the
Hawking temperature from Eq.(1), and for the area the quantity
$$
A_{eff} = 4\pi(3\sqrt{3}{G\over{c^2}})^2\,M^2, \eqno(8)
$$
which is the effective area of the black hole when `light bending' is taken
into account.[26]

            Now, is it possible to infer a similar $M^{-2}$-dependence
for the black hole luminosity from Bekenstein's proposal? One must keep in
mind that the concepts like Hawking temperature and black hole
luminosity refer to observations made by an observer far from, and at
rest with respect to, the black hole. Because of this, we begin
by asking a simple question: What does
such an observer actually observe? The answer is that he observes 
{\it wave packets\/} emanating from the black hole.
Now, Bekenstein's proposal implies that
the energy eigenvalues corresponding to the stationary states of the black
hole are
$$
E_n = \left(\alpha{{\hbar c^5}\over{16\pi G}}\right)^{1\slash
2}\,\sqrt{n},\eqno(9)
$$
where $n$ is an integer. The reason for an emission of a wave packet
is a transition, due to interactions between spacetime and matter fields,
between stationary states. If the black hole undergoes a transition
from a state with energy $E_n$ to the state with energy $E_{n-m}$
($m=1,2,3...$), the energy of the wave packet emitted in this process is
$$
\epsilon_m = m\hbar\omega_0,\eqno(10)
$$
where $\omega_0$ is the fundamental angular frequency of Eq.(5). We 
shall assume that the transitions where $m$ is small are preferred. This
is because black hole radiation, according to the semi-classical
picture, involves creation of virtual particle-antiparticle pairs
near the black hole horizon, and `swallowing' one member of some of the pairs.
The greater the energy $\epsilon$ of a particle of any virtual pair
is, the smaller is the
probability that the pair will `live' long enough
that the hole is able to `swallow' one of the particles. (To be more
quantitative, we note that Hawking's semi-classical theory implies
that the probability of this process is proportional to
$e^{-{\epsilon\slash(kT)}}$, which goes rapidly to zero when 
$\epsilon\gg\hbar\omega_0$.)

           Now, the transition from the state with energy $E_n$ to the
state with energy $E_{n-m}$ takes place, from the point of view of a
distant observer at rest, within a certain finite time interval, and
therefore the length of the emitted wave packet is
finite. For wave packets corresponding to massless particles this
length is, characteristically,
$$
l\sim \tau_m c,\eqno(11)
$$
where $\tau_m$ is the life time, relative to our observer, of the
state with energy $E_n$ before
this state decays into the state with energy $E_{n-m}$. Since the
lengths of the wave packets are finite, the packets
are superpositions of waves
of different frequencies, and therefore of different energies. In
other words, there is an {\it uncertainty\/}
$\Delta\epsilon_m$ in the energy
$\epsilon_m$ of the wave packet when it achieves the
observer. According to Heisenberg's uncertainty principle, we have
$$
\Delta\epsilon_m \sim {{\hbar}\over{\tau_m}}.\eqno(12)
$$
Since we assume that the transitions where $m$ is small are preferred, we
can take this uncertainty, when $m=1$, as an estimate to the
uncertainties $\Delta E_n$ of the energies $E_n$ of the stationary states of
the hole. Moreover, we can estimate the luminosity of the black hole
in terms of the uncertainty $\Delta E_n$. We
obtain
$$
L \sim {{\epsilon_1}\over{\tau_1}} \sim \omega_0\Delta E_n.\eqno(13)
$$
Under the assumption that the transitions with small $m$ are dominant,
one may expect Eq.(13) to be a good estimate to the black hole luminosity.
This is because we can then think of black hole radiation as arising from a
chain of transitions from higher to lower energy states, such that the most
common transition is the decay from the state with energy $E_n$ into a state
with energy $E_{n-1}$. If $n$ and $n'$ are large enough, the
states with energies $E_n$ and $E_{n-1}$ differ from each other in the same
way as the states with energies $E_{n'}$ and $E_{n'-1}$, and so we may
assume that the lifetimes of the energy eigenstates are equal.

            We shall now make a contact with the continuous theory
by assuming that the interaction between spacetime and matter fields
becomes so strong that it overshadows the effects of quantum mechanics
of the spacetime. The stronger the interaction between spacetime and
matter fields, the shorter is the life time of each stationary
state of the black hole. Indeed, we should perhaps no more talk about
stationary but rather about metastable states of the hole. 
According to Heisenberg's uncertainty principle, short life time
implies large uncertainties $\Delta E_n$ in the energies of these
metastable states. When the interaction between spacetime and matter
fields becomes so strong that the uncertainty $\Delta E_n$ is, for
each $n$, of the same order of magnitude as the energy difference
between states with energies $E_n$ and $E_{n-1}$, i. e. if
$$
\Delta E_n \sim \hbar\omega_0\eqno(14)
$$
for each $n$, the energy levels overlap each other, and the continuous
theory is, in effect, recovered. It follows from Eqs.(5), (13) and (14) 
that the black hole luminosity is, in this limit, if we take $\alpha=1$,
$$
L \sim \left({{c^3}\over{32\pi G}}\right)^2\hbar\,{1\over{M^2}}.\eqno(15)
$$
In other words, we have recovered the $M^{-2}$-dependence of the black
hole luminosity. When Eq.(15) is put in numbers, we get
$$
L \sim 10^{33} {{{\rm W}{\rm kg}^2}\over{M^2}},\eqno(16)
$$
which agrees with Page's estimate in Eq.(6).

       In this paper we have formulated a qualitative argument, based on
Heisenberg's uncertainty principle, to the effect that the proposal (2) of
Bekenstein and others for a discrete black hole area spectrum  may in fact
be compatible with Hawking's results of black hole radiation.
More precisely, we argued that if interaction between
spacetime and matter fields is so strong that the uncertainty of
unperturbed stationary energy levels predicted by Bekenstein's
proposal is of the same order of magnitude as the energy
difference between these levels---in which case the continuous
spectrum is, in effect,
recovered---the black hole luminosity is
proportional to the factor $M^{-2}$.
This result is in agreement with
the predictions obtainable from
Hawking's semi-classical theory, which gives the
black-body spectrum for black hole radiation. One may therefore
feel tempted to regard Bekenstein's proposal and Hawking's
semi-classical theory as special cases of the one and the same
prospective theory of
quantum black holes. The former describes black holes in the absence
of matter fields, whereas the latter describes the behavior of matter
fields on the black hole background.
However, it must be emphasized that our arguments have been highly
qualitative. It remains to be seen whether our view on the relationship
between Bekenstein's proposal and Hawking's theory will be validated by a
quantitative theory of quantum spacetime geometry coupled to matter fields.

\bigskip

\bigskip

\centerline{\title Acknowledgments}

\bigskip

       I am most grateful to Dr. Jorma Louko for his constructive
criticism during the preparation of this paper as well as for
suggesting several improvements to the original manuscript. I am also
grateful to DAMTP for hospitality during my visit. This
research was supported by the Finnish Cultural Foundation.

\bigskip

\bigskip

\centerline{\title References}

\bigskip

\bigskip

[1] S. W. Hawking, Commun. Math. Phys. {\bf 43}, 199 (1975).

\medskip

[2] J. D. Bekenstein, Lett. Nuovo Cimento {\bf 11}, 467 (1974).

\medskip

[3] J. D. Bekenstein and V. F. Mukhanov,Phys. Lett. B {\bf 360}, 7 (1995).

(gr-qc/9505012)

\medskip

[4] V. F. Mukhanov, Pis'ma Zh. Eksp. Teor. Fiz. {\bf 44}, 50 (1986)
[JETP Lett.\ 

{\bf 44}, 63 (1986)].

\medskip

[5] I. Kogan, Pis'ma Zh. Eksp. Teor. Fiz. {\bf 44}, 209 (1986)
[JETP Lett. {\bf 44}, 

267 (1986)].

\medskip

[6] P. O. Mazur, Gen.  Relativ. Gravit. {\bf 19}, 1173 (1987).

\medskip

[7] P. O. Mazur, Phys. Rev. Lett. {\bf 57}, 929 (1986);
{\bf 59}, 2380 (1987).

\medskip

[8] V. F. Mukhanov,
in {\it Complexity, Entropy, and the Physics of
Informa-

tion},
SFI Studies in the Sciences of Complexity, Vol.\ III,
edited by W.~H. 

Zurek (Addison--Wesley, New York, 1990).

\medskip

[9] U. H. Danielsson and M. Schiffer, Phys. Rev. D {\bf 48}, 4779 (1993).

\medskip

[10] J.
%%SS Garc\'{i}a--Bellido,
Garc\'{\i}a--Bellido,
``Quantum Black Holes,"
Report SU-ITP-93/4, hep-

th/9302127.

\medskip

[11] Y. Peleg, ``Quantum dust black holes,'' Report BRX-TH-350, hep-th/9307057.

\medskip

[12] M. Maggiore, Nucl. Phys. {\bf B429}, 205 (1994).
(gr-qc/9401027)

\medskip

[13] I. Kogan,
``Black Hole Spectrum, Horison Quantization and All That,"

Report OUTP--94--39P, hep-th/9412232.

\medskip

[14] C. O. Lousto,
Phys. Rev. D {\bf 51}, 1733 (1995).
(gr-qc/9405048)

\medskip

[15] Y. Peleg,
Phys. Lett. B {\bf 356}, 462 (1995).

\medskip

[16] P. O. Mazur,
Acta Phys. Pol.  B {\bf 26}, 1685 (1995).
(hep-th/9602044)

\medskip

[17] P. O. Mazur,
``Gravitation, the Quantum, and Cosmological Constant,"

e-print hep-th/9603014.

\medskip

[18] A.~Barvinsky and
G.~Kunstatter,
``Exact Physical Black Hole States in 

Generic 2-D Dilaton Gravity,"
e-print hep-th/9606134.

\medskip

[19] A. Barvinsky and
G. Kunstatter,
``Mass Spectrum for Black Holes in 

Generic 2-D Dilaton Gravity,"
e-print gr-qc/9607030.

\medskip

[20] H. A. Kastrup,
``On the quantum levels of isolated spherically symmetric 

gravitational
systems,'' Report PITHA 96/16, gr-qc/9605038.

\medskip

[21] J. Louko and J. M\"{a}kel\"{a}, ``Area spectrum of the Schwarzschild
black hole'', 

Report DAMTP 96-53, PP96-98, gr-qc/9605058 (accepted for
publication in 

Phys. Rev. D.)

\medskip

[22] D. N. Page, Phys. Rev. D{\bf 13}, 198 (1976)

\medskip

[23] D. N. Page, Phys. Rev. D{\bf 14}, 3260 (1976)

\medskip

[24] D. N. Page, Phys. Rev. D{\bf 16}, 2402 (1977)

\medskip

[25] N. D. Birrell, P. C. Davies , in {\it Quantum Fields in Curved
Space} (Cam-

bridge University Press, Cambridge 1982)

\medskip

[26] R. M. Wald, in {\it Quantum Field Theory in Curved Spacetime and
Black 

Hole Thermodynamics} (The University of Chicago Press, Chicago
1994)

\medskip

\bye